\documentclass[aps, prd, twocolumn, lengthcheck, superscriptaddress, showpacs, letterpaper, nofootinbib]{revtex4-1}

\usepackage{amsfonts}
\usepackage{amsmath}
\usepackage{amssymb}
\usepackage{xcolor}
\usepackage{graphicx}
\usepackage{subfigure}
\usepackage{slashed}
\usepackage{hyperref}
\usepackage{amsthm}
\usepackage{mathrsfs}
\usepackage{color}
\usepackage{epsfig}
\usepackage{epstopdf}
\usepackage{bm}

\def\be{\begin{eqnarray}}\def\ee{\end{eqnarray}}

\def\lsim{\mathrel{\rlap{\lower3pt\hbox{\hskip1pt$\sim$}}
		\raise1pt\hbox{$<$}}} %less than or approx. symbol
\def\gsim{\mathrel{\rlap{\lower3pt\hbox{\hskip1pt$\sim$}}
		\raise1pt\hbox{$>$}}} %greater than or approx. symbol

\allowdisplaybreaks

\begin{document}

\title{Vector meson effects on the multi-skyrmion states from the rational map ansatz}

\author{Jun-Shuai Wang}
\affiliation{College of Physics, Jilin University, Changchun, 130012, China}

\affiliation{School of Fundamental Physics and Mathematical Sciences,
	Hangzhou Institute for Advanced Study, UCAS, Hangzhou, 310024, China}

\author{Yong-Liang Ma}
\email{ylma@ucas.ac.cn}
%\affiliation{College of Physics, Jilin University, Changchun, 130012, China}
\affiliation{School of Fundamental Physics and Mathematical Sciences,
	Hangzhou Institute for Advanced Study, UCAS, Hangzhou, 310024, China}

\begin{abstract}

The roles of the lightest vector mesons $\rho$ and $\omega$ in the multi-skyrmion states are studied using the hidden local symmetry approach upto the next to leading order including the homogeneous Wess-Zumino terms. The low energy constants in the effective field theory are determined by using the Sakai-Sugimoto model and the flat-space five-dimensional Yang-Mills action. With only two inputs, $m_\rho$, and $f_\pi$, all the low energy constants can be determined without ambiguity. The vector meson effects can be investigated by integrating them in order and the influence from the geometry can be clarified by comparing the results using the low energy constants estimated from the Sakai-Sugimoto model and the flat-space five-dimensional Yang-Mills action. We find that the $\rho$ meson reduces the
masses of the multi-skyrmion states and increases the overlaps of the constituents of the multi-skyrmion states while the $\omega$ meson repulses the constituents of the multi-skyrmion states and increases their masses, therefore these vector mesons are important in Skyrme model approach to nuclei. we also find that the warping factor which is an essential element in the holographic model of QCD affects the properties of the multi-skyrmion states and cannot be ignored.

\end{abstract}

\maketitle

\section{Introduction}

The Skyrme model~\cite{Skyrme:1961vq,Skyrme:1962vh} as a nonlinear theory of mesons anchored on the chiral symmetry breaking of QCD provides a unified framework to study the single baryons, multi-baryon states and nuclear matter~\cite{RhoZahed,Zahed:1986qz,Ma:2016gdd,Ma:2019ery,Manton:2022fcb} when the skyrmions are regarded as baryons in the limit of large $N_c$~\cite{Witten:1979kh,Witten:1983tx,Witten:1983tw}.

In the skyrmion approach to nuclear physics, it is found that the vector mesons play indispensible roles~\cite{	Zahed:1984qv,Meissner:1986ka,Igarashi:1985et}. With the help of effective models in higher dimensions, the vector meson effects can be studied without ambiguity. In Refs.~\cite{Nawa:2006gv,Ma:2012kb,Ma:2012zm}, the authors studied the skyrmion properties by dimensionally deconstructing a five-dimensional holographic model, the Sakai-Sugimoto model~\cite{Sakai:2004cn,Sakai:2005yt}, to an effective theory of vector mesons in four dimensions, the hidden local symmetry (HLS) approach~\cite{Bando:1984ej,Bando:1987br, Harada:2003jx}. In this approach, all the low energy constants (LECs) can be fixed with only two inputs $f_\pi$ and $m_\rho$ except the parameter $a$ which is proved that any physical quantities calculated with the HLS Lagrangian induced from hQCD models are actually independent of it~\cite{Ma:2012zm}. It is found that the inclusion of the $\rho$ meson reduces the soliton mass, which makes the Skyrmion come closer to the Bogomol’nyi-Prasad-Sommerfield soliton, but the role of the $\omega$ meson is found to increase the soliton mass. Similarly, using a $(4+1)$-dimensional Yang-Mills theory which may be written as a $(3+1)$-dimensional BPS Skyrme model, the iso-vector hadron resonances $\rho$ and $a_1$ are found suppress the skyrmion mass and the more resonances included the further the suppression~\cite{Sutcliffe:2010et,Sutcliffe:2011ig}.

The skyrmion approach to the multi-skyrmion states is achieved by using the product ansatz~\cite{Skyrme:1961vq,Battye:1996nt,Gudnason:2018aej} or the rational map ansatz~\cite{Houghton:1997kg,Krusch:2002by,Gudnason:2018jia} since the multi-skyrmion states obtained by naively extending the boundary conditions of hedgehog ansatz are unstable~\cite{Weigel:1986zc}. A generic property of the multi-skyrmion states is that their shapes are not spherical like the $B=1$ skyrmion but have special symmetries. Moreover, it is found that the states with large baryon numbers have hollow structure in the chiral limit and this hollow structure may be unstable when the physical pion mass is considered (see, e.g. Ref.~\cite{Battye:2006na} for a review.). This hollow structure is interesting for understanding the possible multi-layer structure of neutron stars~\cite{Nelmes:2011zz} considering that skyrmion matter at high density has the sheet structure made of half-skyrmions~\cite{Park:2019bmi}

In terms of the standard Skyrme model, the multi-skyrmion states were investigated using the rational map ansatz~\cite{Houghton:1997kg}. It is found that some states are not bound. However, another approach using  different numerical algorithms does not find these unbounded states~\cite{Battye:1997qq}. When the Skyrme model is extended to include the positive pion mass, the structure of the multi-skyrmion states are changed ~\cite{Battye:2004rw} and the $\alpha$-cluster structure of nuclei is found~\cite{Battye:2006na}. Moreover, when the Skyrme model is extended to include the vector meson $\rho$ using an effective (3+1)-dimensional BPS theory truncated from a (4+1)-dimensional Yang-Mills theory, the masses of the multi-skyrmion states with baryon number upto $B=4$ are found to be suppressed~\cite{Sutcliffe:2011ig,Naya:2018mpt}. In the same framework including massive pions, people found that the clustering structure of the light nuclei can be yielded and the binding energies are very close to the nuclear data~\cite{Naya:2018kyi}.

Although many accesses to the multi-skyrmion states have been done in the literature, there are still some, for example, the following ambiguities are waiting for clarification: What is the effect of the iso-scalar vector meson $\omega$ which is responsible for repulsive force in nuclear physics on the multi-baryon states with baryon number $B>2$? Whether the characters of the multi-skyrmion states are changed when the $\rho$ is considered as an independent degree of freedom or integrated out from the theory such that its effect is hidden in the Skyrme parameter? What is the influence of the geometry in the five-dimension which affects the values of the LECs?  We will clarify these ambiguities in this work systematically using the rational map ansatz and leave the discussion of the structures of the multi-skyrmion states at global minima to future work.

We use the hidden local symmetry approach for the vector mesons in (3+1)-dimension which is developed in the nonlinear realization of chiral symmetry~\cite{Bando:1984ej,Bando:1987br, Harada:2003jx}. The Lagrangian is considered up to the next to leading order including the homogeneous Wess-Zumino (hWZ) terms which are responsible for the omega meson effect. To control the ambiguities for the LECs, we resort to the effective models in (4+1)-dimension, i.e., the Sakai-Sugimoto (SS) model and the (4+1)-dimensional Yang-Mills theory (BPS model). Comparing the results from the HLS with the LECs fixed using a certain effective model, we can check the resonance effects by integrating the resonance in order. Moreover, the distinction between the results obtained using the LECs yielded from the SS model and BPS model tells us the effect of the warping factor in the five-dimension.

The rest of this work is arranged as follows: In Sec.~\ref{sec:HLS} we outline the effective field theory that will be used in this work and rational map ansatz up to baryon number $B=8$. In Sec.~\ref{sec:Num} we list our numerical results and compare these values obtained from different models. Our conclusion and discussion are given in Sec.~\ref{sec:conclusion}. The expression of the masse of the multi-skyrmion state is given in Appendix~\ref{sec:app}.

\section{The hidden local symmetry approach for vector mesons}

\label{sec:HLS}

To see the effects of the vector mesons rho and omega on the multi-skyrmion states, among a variety of effective approaches we use the hidden local symmetry (HLS) method to include these vector mesons in the chiral effective theory~\cite{Bando:1984ej, Bando:1987br,Harada:2003jx}. We consider the HLS up to the next to leading order including the homogeneous Wess-Zumino (hWZ) terms which are responsible for the contribution from omega meson.

The full symmetry considered in this work is $ G_{\rm full} =  [SU(2)_{L} \times SU(2)_{R}]_{\rm chiral} \times [ U(2)_{V}]_{\rm HLS}$ with $ [U(2)_{V}]_{\rm HLS}$ being the HLS. The HLS Lagrangian with symmetry $ G_{\rm full}$ can be written in terms of the Maurer-Cartan 1-forms
\be
\hat{\alpha}_{\perp \mu} & = & \frac{1}{2i}\left(D_\mu \xi_R\cdot \xi_R^\dagger - D_\mu \xi_L\cdot \xi_L^\dagger\right),\nonumber\\
\hat{\alpha}_{\parallel \mu} & = & \frac{1}{2i}\left(D_\mu \xi_R\cdot \xi_R^\dagger + D_\mu \xi_L\cdot \xi_L^\dagger\right),
\ee
with the chiral fields $\xi_{L,R}$ which in the unitary gauge are written as
\be
\xi_L^\dagger = \xi_R \equiv e^{i\pi/2f_\pi}
\ee 
where $\pi = \bm{\pi}\cdot \bm{\tau}$ with $\bm{\tau}$ being the Pauli matrices. Without considering the external sources, we write the covariant derivative as
\be
D_\mu \xi_{L,R} = \left(\partial_\mu - i V_\mu\right)\xi_{L,R},
\ee
with $V_\mu$ being the gauge boson of the HLS. After breaking the HLS and in the unitary gauge, the field $V_\mu$ is expressed in terms of the vector meson fields as 
\be
V_\mu & = & \frac{g}{2}\left(\omega_\mu + \rho_\mu \right),
\ee 
with 
\be
\rho_\mu & = & \bm{\rho}_\mu \cdot \bm{\tau} = 
\left(\begin{array}{cc}
\rho_\mu^0	& \sqrt{2}\rho_\mu^+ \\
\sqrt{2}\rho_\mu^-	& {}-\rho_\mu^0
\end{array}
\right)
\ee
In addition to the two Maurer-Cartan 1-forms $\hat{\alpha}_{\perp,\parallel, \mu}$, due to the gauge field of the HLS, the third block in the construction of the HLS Lagrangian is the field strength tensor 
\be
V_{\mu\nu} = \partial_\mu V_\nu - \partial_\nu V_\mu - i \left[V_\mu,V_\nu\right].
\ee

With the above discussion, one can construct the HLS Lagrangian which will be used in this work up to $O(p^4)$ as~\cite{Harada:2003jx}
\be
\mathcal{L} & = & \mathcal{L}_{(2)} + \mathcal{L}_{(4)} + \mathcal{L}_{\rm anom} .
\label{eq:HLSall}
\ee
The leading order Lagrangian, the $O(p^2)$ terms, $\mathcal{L}_{(2)}$ in the chiral limit which will be considered in this work is
\be
\mathcal{L}_{(2)} & = & f_{\pi}^2 \text{Tr}( \hat{\alpha}_{\perp\mu} \hat{\alpha}_{\perp}^{\mu} )
        + a f_{\pi}^2 \text{Tr}( \hat{\alpha}_{\parallel\mu} \hat{\alpha}_{\parallel}^{\mu} )\nonumber\\
        & &{} - \frac{1}{2g^2} \text{Tr}( V_{\mu\nu} V^{\mu\nu} ),
\ee
where $f_\pi$ is the pion decay constant, $a$ is the parameter of
the HLS, $g$ is the coupling constant of the hidden local gauge field---the vector meson field. For the  $O(p^4)$ Lagrangian, we only consider the terms having one trace since the terms including two traces are suppressed by $1/N_c$. Then the $O(p^4)$ Lagrangian we will use is given by
\be
\mathcal{L}_{(4)} & = & \mathcal{L}_{(4)y} + \mathcal{L}_{(4)z},
\ee
where
\be
    \mathcal{L}_{(4)y} & = & y_1 \text{Tr}[ \hat{\alpha}_{\perp\mu} \hat{\alpha}_{\perp}^{\mu} \hat{\alpha}_{\perp\nu} \hat{\alpha}_{\perp}^{\nu} ]
        + y_2 \text{Tr}[ \hat{\alpha}_{\perp\mu} \hat{\alpha}_{\perp\nu} \hat{\alpha}_{\perp}^{\mu} \hat{\alpha}_{\perp}^{\nu} ] \nonumber \\
        & &{} + y_3 \text{Tr}[ \hat{\alpha}_{\parallel\mu} \hat{\alpha}_{\parallel}^{\mu} \hat{\alpha}_{\parallel\nu} \hat{\alpha}_{\parallel}^{\nu}]
        + y_4 \text{Tr}[ \hat{\alpha}_{\parallel\mu} \hat{\alpha}_{\parallel\nu} \hat{\alpha}_{\parallel}^{\mu} \hat{\alpha}_{\parallel}^{\nu}]  \nonumber \\
        & &{} + y_5 \text{Tr}[ \hat{\alpha}_{\perp\mu} \hat{\alpha}_{\perp}^{\mu} \hat{\alpha}_{\parallel\nu} \hat{\alpha}_{\parallel}^{\nu}]
        + y_6 \text{Tr}[ \hat{\alpha}_{\perp\mu} \hat{\alpha}_{\perp\nu} \hat{\alpha}_{\parallel}^{\mu} \hat{\alpha}_{\parallel}^{\nu}]
        \nonumber \\
        & &{} + y_7 \text{Tr}[ \hat{\alpha}_{\perp\mu} \hat{\alpha}_{\perp\nu} \hat{\alpha}_{\parallel}^{\nu} \hat{\alpha}_{\parallel}^{\mu}]\nonumber \\
        & &{} + y_8 \{ \text{Tr}[ \hat{\alpha}_{\perp\mu} \hat{\alpha}_{\parallel}^{\mu} \hat{\alpha}_{\perp\nu} \hat{\alpha}_{\parallel}^{\nu} ]
        + \text{Tr}[ \hat{\alpha}_{\perp\mu} \hat{\alpha}_{\parallel\nu} \hat{\alpha}_{\perp}^{\nu} \hat{\alpha}_{\parallel}^{\mu} ] \}\nonumber\\
        & &{} + y_9 \text{Tr}[ \hat{\alpha}_{\perp\mu} \hat{\alpha}_{\parallel\nu} \hat{\alpha}_{\perp}^{\mu} \hat{\alpha}_{\parallel}^{\nu} ], \\
    \mathcal{L}_{(4)z} & = & i z_4 \text{Tr}[ V_{\mu\nu} \hat{\alpha}_{\perp}^{\mu} \hat{\alpha}_{\perp}^{\nu} ]
        + i z_5 \text{Tr}[ V_{\mu\nu} \hat{\alpha}_{\parallel}^{\nu} \hat{\alpha}_{\parallel}^{\nu}].
\ee
For the anomalous parity part, the Lagrangian $\mathcal{L}_{\rm anom}$ has expression
\be
\mathcal{L}_{\rm anom} & = & \frac{N_c}{16\pi^2} \sum_{i=1}^{3} C_i \mathcal{L}_i,
\ee
where
\begin{subequations}
\be
\mathcal{L}_{1} & = & i \text{Tr} [ \hat{\alpha}_L ^3 \hat{\alpha}_R - \hat{\alpha}_R ^3 \hat{\alpha}_L ], \\
\mathcal{L}_{2} & = & i \text{Tr} [ \hat{\alpha}_L \hat{\alpha}_R \hat{\alpha}_L \hat{\alpha}_R ], \\
\mathcal{L}_{3} & = & \text{Tr} [ F_V (\hat{\alpha}_L \hat{\alpha}_R - \hat{\alpha}_R \hat{\alpha}_L) ],
\ee
\label{eq:Lanom}
\end{subequations}
with the 1-form and 2-form fields
\be
\hat{\alpha}_L & = & \hat{\alpha}_{\parallel} - \hat{\alpha}_{\perp},\quad \hat{\alpha}_R = \hat{\alpha}_{\parallel} + \hat{\alpha}_{\perp},\nonumber\\
F_V & = & dV - iV^2.
\ee

To study the properties of the multi-skyrmion states using the Lagrangian~\eqref{eq:HLSall} from the rational map ansatz, we parameterize the chiral field as~\cite{Houghton:1997kg},
\be
\xi(r) & = & \text{exp} \left [ i \bm{\tau} \cdot \hat{\bm{n} } \frac{F(r)}{2} \right ]
\ee
where
\be
\hat{\bm{n} } & = & \frac{ 1 }{ 1+\lvert R \rvert ^2  } ( 2 \text{Re}(R), 2\text{Im}(R), 1-\lvert R \rvert ^2 )
\label{eq:ansatzpion}
\ee
with $R$ being the rational map which is a function of the complex coordinate $z$ on a Riemann unit two-sphere and $r$ as the distance from the origin. For a baryon number $B$ state, the rational map $R(z) = p/q$ with $p$ and $q$ are polynominal in $z$ that $max[deg(p, deg(q))]=N$ and $p$ and $q$ have no common factors. Explicitly, for $B=1, 2, \cdots, 8$, $R(z)$ has the following form~\cite{Houghton:1997kg}
\begin{description}
	\item[$N=1$] $R(z)=z$, the hadgehog map.
	\item[$N=2$] $R(z)=\frac{z^2-a}{-az^2+1}$,  with $a$ being a real parameter and $-1\leq a \leq 1$.
	\item[$N=3$] $R(z)=\frac{\sqrt{3}az^2-1}{z(z^2-\sqrt{3}a)}$, with $a$ being a complex parameter.
	\item[$N=4$] $R(z)=c\frac{z^4+2\sqrt{3}iz^2+1}{z^4-2\sqrt{3}iz^2+1}$, with $c$ being a real parameter.
	\item[$N=5$] $R(z)=\frac{z(z^4+bz^2+a)}{az^4-bz^2+1}$, with $a$ and $b$ as real parameters.
	\item[$N=6$] $R(z)=\frac{z^4+ia}{z^2(iaz^4+1)}$, with $a$ being a real parameter .
	\item[$N=7$] $R(z)=\frac{bz^6-7z^4-bz^2-1}{z(z^6+bz^4+7z^2-b)}$, with $b$ being a complex parameter.
	\item[$N=8$] $R(z)=\frac{z^6-a}{z^2(az^6+1)}$, with $a$ being a real parameter .
\end{description}
For the vector mesons $\rho$ and $\omega$, we apply the following configurations~\cite{Meissner:1986js,Imai:1989jh},
\be
\omega_{\mu} & = & W(r) \delta_{0\mu}, \nonumber\\
\rho_0 & = & 0,\nonumber\\
\rho_{i} & = &{} - \frac{G(r)}{g} \bm{\tau} \cdot ( \hat{\bm{n}} \times \partial_{i}{ \hat{\bm{n}} } ).
\label{eq:ansatzV}
\ee
The profiles of the meson fields satisfy the following boundary conditions,
\be
F(0) & = & \pi,\;\;\;\; F(\infty) = 0, \nonumber\\
G(0) & = & 2,\;\;\;\; G(\infty) = 0, \nonumber\\
W'(0) & = & 0,\;\;\;\; W(\infty) = 0 .
\label{eq:BCs}
\ee

By using the effective Lagrangian~\eqref{eq:HLSall} and the antsatz~\eqref{eq:ansatzpion} and \eqref{eq:ansatzV} one can easy derive the expression of the mass of the multi-skyrmion state. The explicit formula is given in Appendix~\ref{sec:app}. Minimizing the mass of the multi-skyrmion state subjecting to the boundary conditions~\eqref{eq:BCs} one can obtain the profiles of $F(r), G(r)$ and $W(r)$ and therefore the mass of the multi-skyrmion states once the LECs are given.

\section{Numerical results for the multi-skyrmion states}

\label{sec:Num}

\subsection{The model and low energy constants}

In order to calculate the properties of the multi-skyrmion states expressed in Appendix~\ref{sec:app}, we should first know the values of the LECs. As one can easily see, there are 18 parameters in Lagrangian~\eqref{eq:HLSall}, $f_\pi, a, g, y_1, y_2, \cdots, y_9, z_4, z_5, c_1, c_2, c_3$ and the vector meson mass $m_V=m_\rho \simeq m_\omega$. Since the mass of the vector mason satisfies the relation
\be
m_V^2 & = & a g^2 f_\pi^2,
\ee
and the empirical values of $f_\pi$ and $m_V = m_\rho \simeq m_\omega$ are well known, 15 parameters are left. We cannot estimate the values of these parameters without ambiguity so far.

Therefore, to finalize the numerical calculation, we estimate the low energy constants from the dual models of QCD in five dimensions, explicitly, the holographic model of QCD from the top-down approach--- the Sakai-Sugimoto (SS) model~\cite{Sakai:2004cn}. As a comparison, we also estimate the LECs by using the Bogomol’nyi-Prasad-Sommerfield (BPS) model~\cite{Sutcliffe:2008sk}---the Yang-Mills theory in five dimensions, for the purpose to show the effect from geometry. We denote the HLS with parameters determined from the SS model as HLS$_{\rm SS}$ and that from the BPS model as HLS$_{\rm BPS}$. Due to the special structure of the 5D Dirac-Born-Infeld (DBI) part of the SS model and the gauge invariance of the (4+1)-dimensional Yang-Mills theory, the LECs in HLS have the following relations
\be
y_1 & = &{}- y_2, \quad y_3={}-y_4, \nonumber\\ 
y_5 & = & 2y_8 = {}-y_9, \quad y_6 ={}-(y_5+y_7)
\ee 
and the omega meson effect only enters through the hWZ terms. In the five-dimensional models, one can prove that the parameter $a$ is related to the normalization of the eigenfunction of the vector mode and the physical quantities calculated with the HLS induced from the five-dimensional models are independent of the parameter $a$ although the values of the LECs depend on it~\cite{Ma:2012zm}. With the choice $a=2$ which reproduces the Kawarabayashi-Suzuki-Riazzudin-Fayyazudin (KSRF) relation and the rho meson dominance in the pion electromagnetic form factor, we list the values of the LECs estimated from the SS model and the BPS model in Table~\ref{table:LECs}~\cite{Ma:2012zm}.  

%%%%%%%%%%%          Table 1          %%%%%%%%%%%%%%%%%%%
\begin{table*}[htbp]
	\caption{\label{table:LECs} Low energy constants of the HLS Lagrangian
		at $O(p^4)$ with $a=2$.}
	\begin{ruledtabular}
		\centering
		\begin{tabular}{clllllllll}
			Model
			& \qquad $y_1$ &\qquad $y_3$ & \qquad $y_5$ & \qquad $y_6$ & \qquad $z_4$ &
			\qquad $z_5$ & \qquad $c_1$ & \qquad $c_2$ & \qquad $c_3$ \\
			\hline
			SS model & $- 0.001096$ & $ - 0.002830$ & $-0.015917$ & $+0.013712$ &
			$0.010795$ & $-0.007325$ & $+0.381653$ & $-0.129602$ & $0.767374$ \\
			BPS model & $- 0.071910$ & $-0.153511$ & $-0.012286$ & $-0.196545$ &
			$0.090338$ & $-0.130778$ & $-0.206992$ & $+3.031734$ & $1.470210$ \\
		\end{tabular}
	\end{ruledtabular}
\end{table*}
%%%%%%%%%%%%%%%%%%%%%%%%%%%%%%%%%%%%%%%%%%%%%%

In the following, for the purpose to investigate the resonance effects on the multi-skyrmion states, we consider three versions of HLS
\begin{description}
	\item[ HLS($\pi,\rho,\omega$)] The HLS with all the $\pi, \rho$ and $\omega$ fields.
	\item[ HLS($\pi,\rho$)]  The HLS with $\pi$ and $\rho$ fields which is obtained by integrating out $\omega$ field from, or equivalently dropping the hWZ terms in, the Lagrangian~\eqref{eq:HLSall}.
	\item[ HLS($\pi$)]  The HLS with only $\pi$ which is obtained from HLS($\pi,\rho,\omega$) by integerating out both $\rho$ and $\omega$ fields. In this scenario, the skyrmion parameter receives contribution from the $y_1, y_2$, and $z_4$ terms in addition to the kinetic term of the vector mesons~\cite{Ma:2012zm}. In this case, the Skyrme parameter $e =7.31$ in HLS$_{\rm SS}(\pi)$ and $e =10.02$ in HLS$_{\rm BPS}(\pi)$.
\end{description}

\subsection{Numerical results}

Equipped with the above estimated LECs, we are ready to calculate the multi-skyrmion states now. In order to obtain the profile functions, we use the finite element method~\cite{buelerPETScPartialDifferential2020} to minimize the total static energy of the system given in Appendix~\ref{sec:app} with respect to the boundary conditions \eqref{eq:BCs}. The advantage of the finite element method is that only the first-order ordinary differential equations (ODEs) need to be handled and if one wants to add more resonances, it is not necessary to derive the tedious equation of motion (second-order ODEs) again.

\subsubsection{Effects from the hadron resonances}

We first consider the properties of the multi-skyrmion states using the HLS with the LECs determined from the SS model to investigate the resonance effects. The masses of the multi-skyrmion states for baryon numbers $B=1,2, \cdots, 8$ are given in Table~\ref{tab:MassSS}.  

% block for baryon mass in ss model
\begin{table}[htp]\centering
	\caption{Masses of the multi-skyrmion states in HLS$_{\rm SS}$ (in unit of $ 4 \pi f_{\pi}^2/m_{\rho} $). Only the hadron degrees of freedom are explicitly written for simplicity.}
	\begin{tabular}{ccccccccc} \hline\hline
		B                   & 1      & 2     & 3     & 4     & 5     & 6     & 7     & 8 \cr \hline
		$(\pi,\rho,\omega)$ & 8.59   & 16.90 & 24.94 & 32.44 & 40.58 & 48.37 & 55.56 & 63.71 \cr  
		$(\pi,\rho)$       & 6.04   & 12.37 & 18.44 & 23.65 & 30.04 & 35.83 & 40.64 & 47.07 \cr
		$(\pi)$    & 6.67  & 13.08 & 19.23 & 24.59 & 31.03 & 36.92 & 41.93 & 48.39 \cr
		\hline\hline
	\end{tabular}
\label{tab:MassSS}
\end{table} 

Comparing the results from HLS$_{\rm SS}(\pi)$ and HLS$_{\rm SS}(\pi,\rho)$ one can see that due to the attractive force from the rho meson, the masses of the multi-skyrmion states are reduced. This is consistent with the experience from the understanding of the nuclear force and the calculation of the skyrmion spectrum~\cite{Ma:2012kb,Ma:2012zm}. However, the comparison of the results from  HLS$_{\rm SS}(\pi,\rho)$ and HLS$_{\rm SS}(\pi,\rho,\omega)$ tell us that, the same as what happens in the skyrmion case~\cite{Ma:2012kb,Ma:2012zm}, due to the repulsive force arising from the omega meson, the masses of the multi-skyrmion states are increased. 

It is interesting to note that, in contrast to \cite{Houghton:1997kg}, in HLS$_{\rm SS}(\pi)$ all the states are bound ones due to the contribution from the higher order terms of HLS. When the $\rho$ is included as an explicit degree of freedom in HLS$_{\rm SS}(\pi,\rho)$ the $B=2$ and $3$ states are not bound since their masses are larger than twice and three times of that of the single skyrmion state, respectively. This is because, when only the $\pi$ and $\rho$ mesons are included in HLS, the model is very close to the Bogomol’ny bound and the force is very weak, as shown in Table~\ref{tab:MassSS}. This can be seen more clearly in the skyrmion crystal approach to nuclear matter~\cite{Ma:2013ooa} and the HLS$_{\rm BPS} (\pi,\rho)$ results which will be shown later. However, when the omega meson, the flavor partner of the rho meson, is considered, all the multi-skyrmion states for $B \ge 2$ are bound and the binding energies are bigger than the corresponding ones in HLS$_{\rm SS}(\pi,\rho)$. This again shows the significance of the omega force in nuclear physics.

\begin{figure*}[htbp] \centering
	\includegraphics[width=1\linewidth]{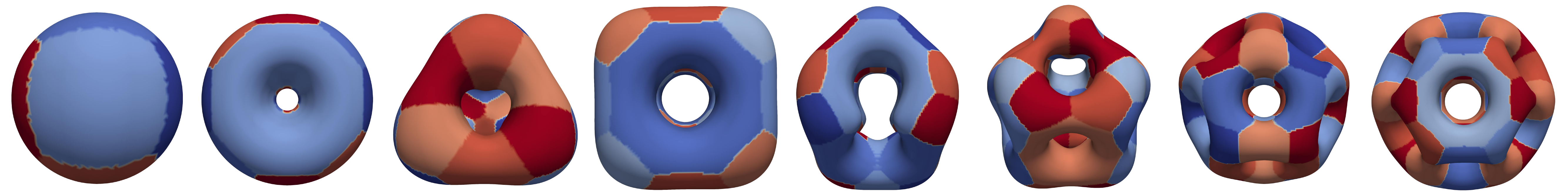}	
	\includegraphics[width=1\linewidth]{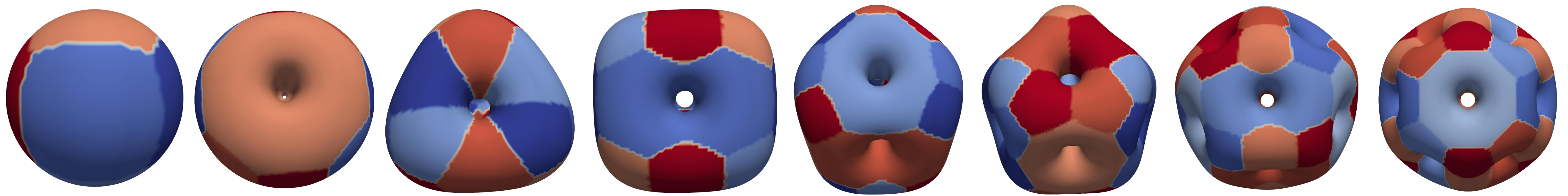}
	\includegraphics[width=1\linewidth]{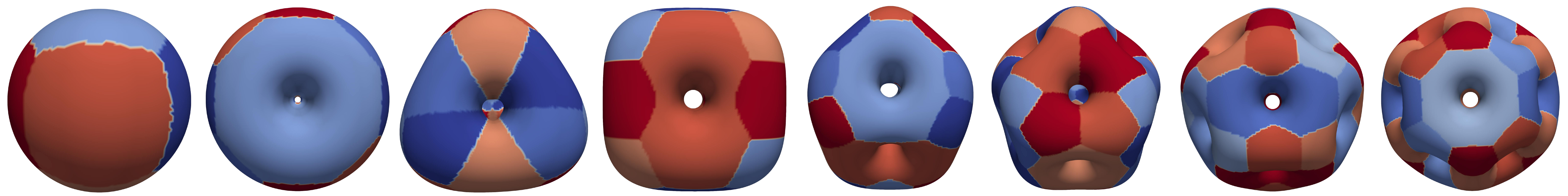}
	\caption{Contour surface with $\mathcal{B}^0=0.01$ in the $\text{HLS}_{\rm SS}(\pi,\rho,\omega)$(upper),  $\text{HLS}_{\rm SS}(\pi,\rho)$ (middle) and  $\text{HLS}_{\rm SS}(\pi)$ (lower) for $B=1,2, \cdots, 8$ (from left to right).} 
	\label{fig:contourSS}
\end{figure*}

To have a deeper understanding of the effects of the hadron resonances, we plot the contour surfaces of the multi-skyrmion states with baryon number density  $\mathcal{B}^0=0.01$ in Fig.~\ref{fig:contourSS}. Comparing the contours from $\text{HLS}_{\rm SS}(\pi)$  and that from $\text{HLS}_{\rm SS}(\pi,\rho)$  one sees that the force from rho meson attracts the constituents of the multi-skyrmion states closer, although the difference is tiny. However, due to the repulsive force from omega meson, the overlap among the constituents in a multi-skyrmion state from $\text{HLS}_{\rm SS}(\pi,\rho,\omega)$ is much smaller than others and the omega meson effect is more significant. 

%\begin{widetext}
%\end{widetext}

\begin{figure}[htbp]
	\centering
	\includegraphics[width=7.0cm]{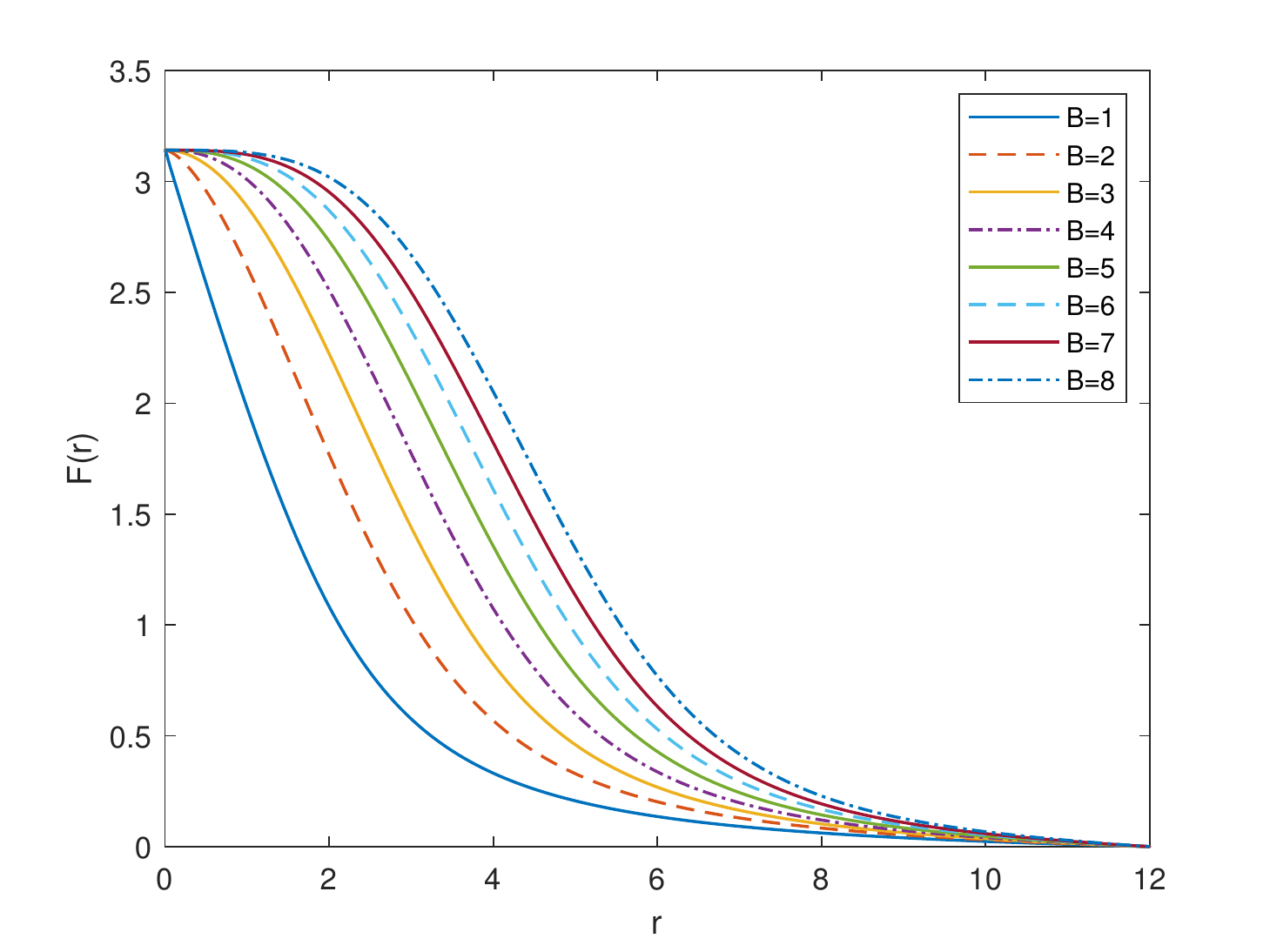}
	\includegraphics[width=7.0cm]{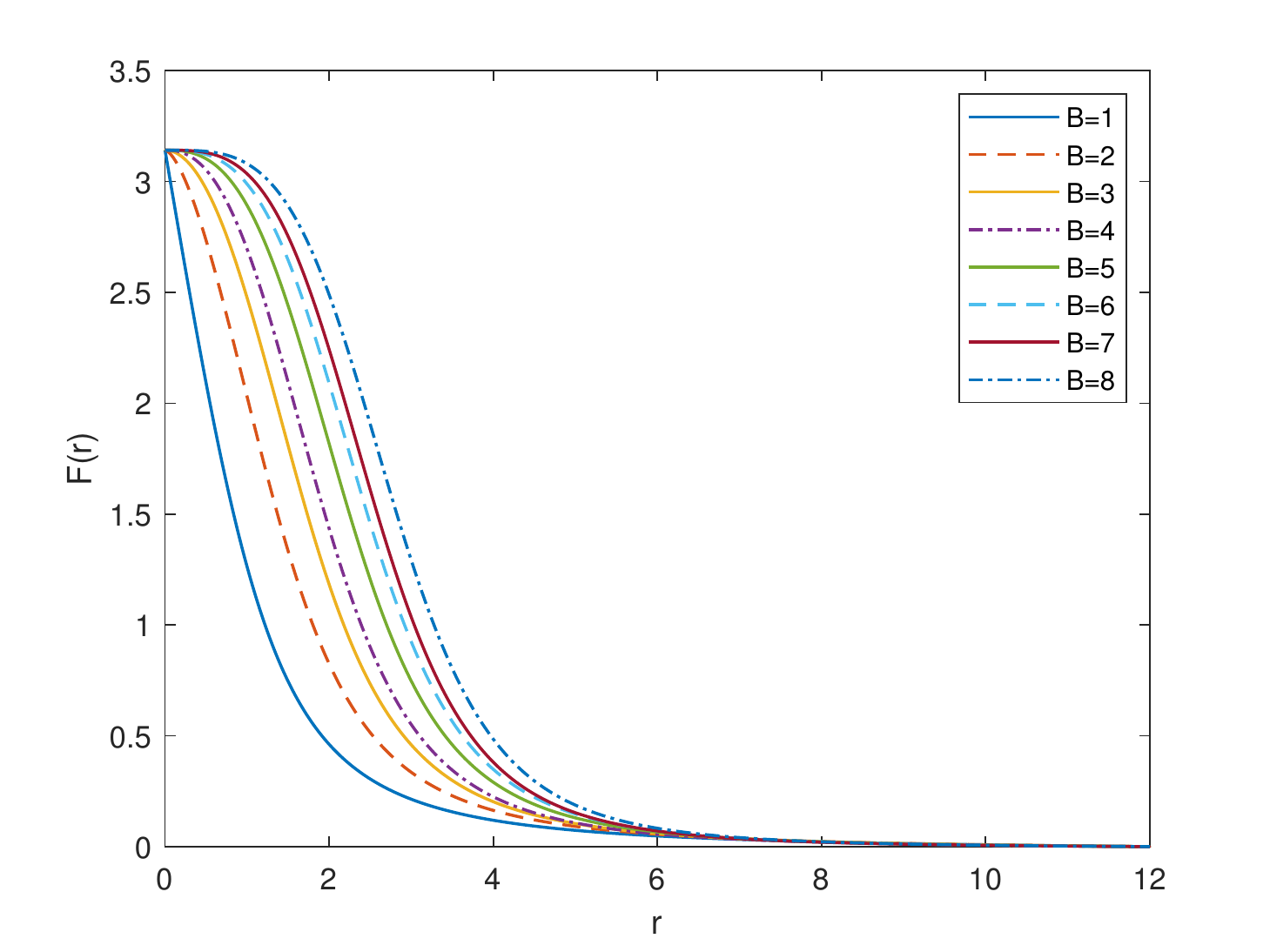}
	\includegraphics[width=7.0cm]{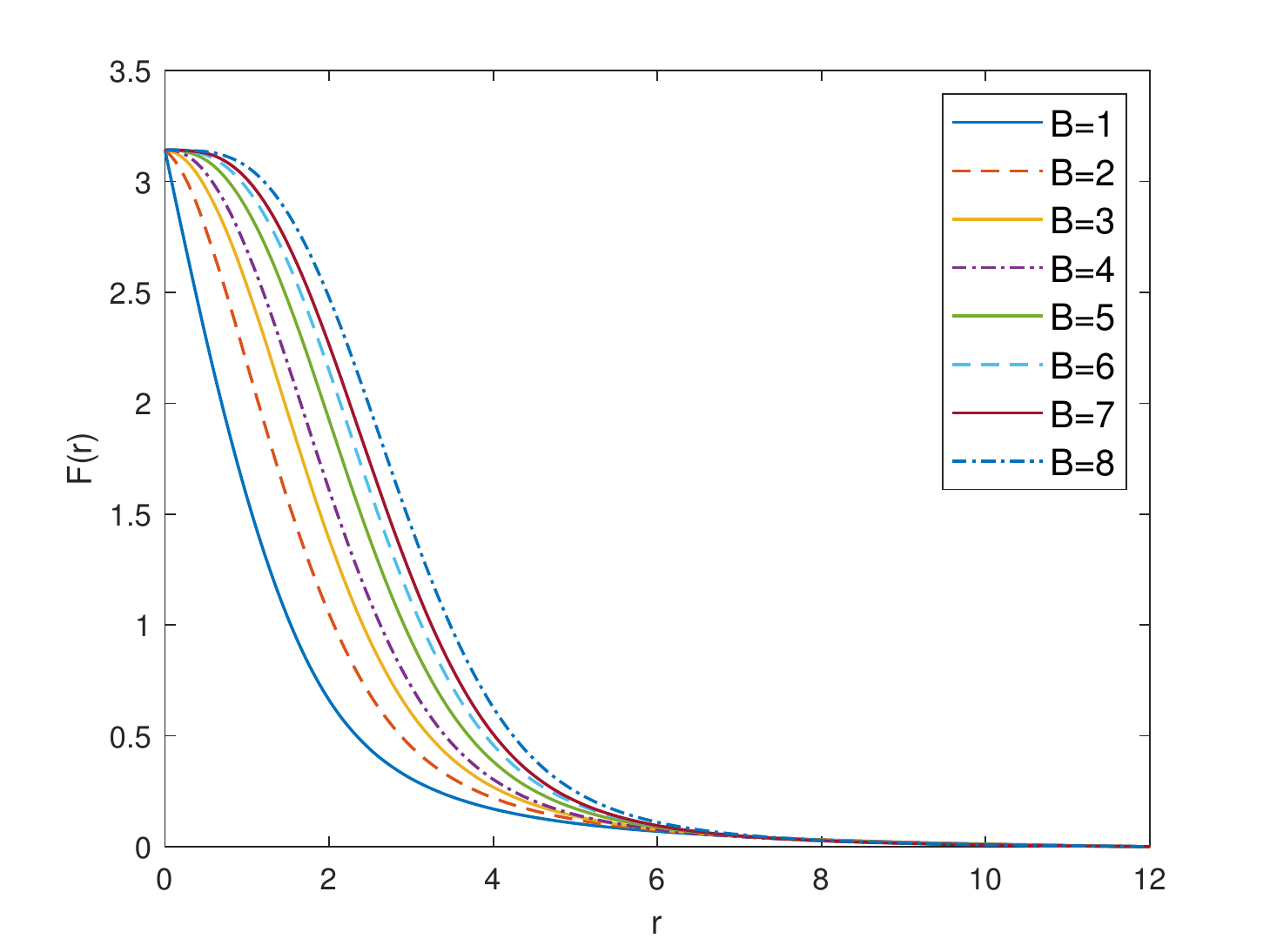}
	\caption{Profile function $F(r)$ in HLS$_{\rm SS}(\pi,\rho,\omega)$ (upper), HLS$_{\rm SS}(\pi,\rho)$ (middle)and HLS$_{\rm SS}(\pi)$ (lower).}
	\label{fig:FSS}
\end{figure}

\begin{figure}[htbp]
	\centering
	\includegraphics[width=7.0cm]{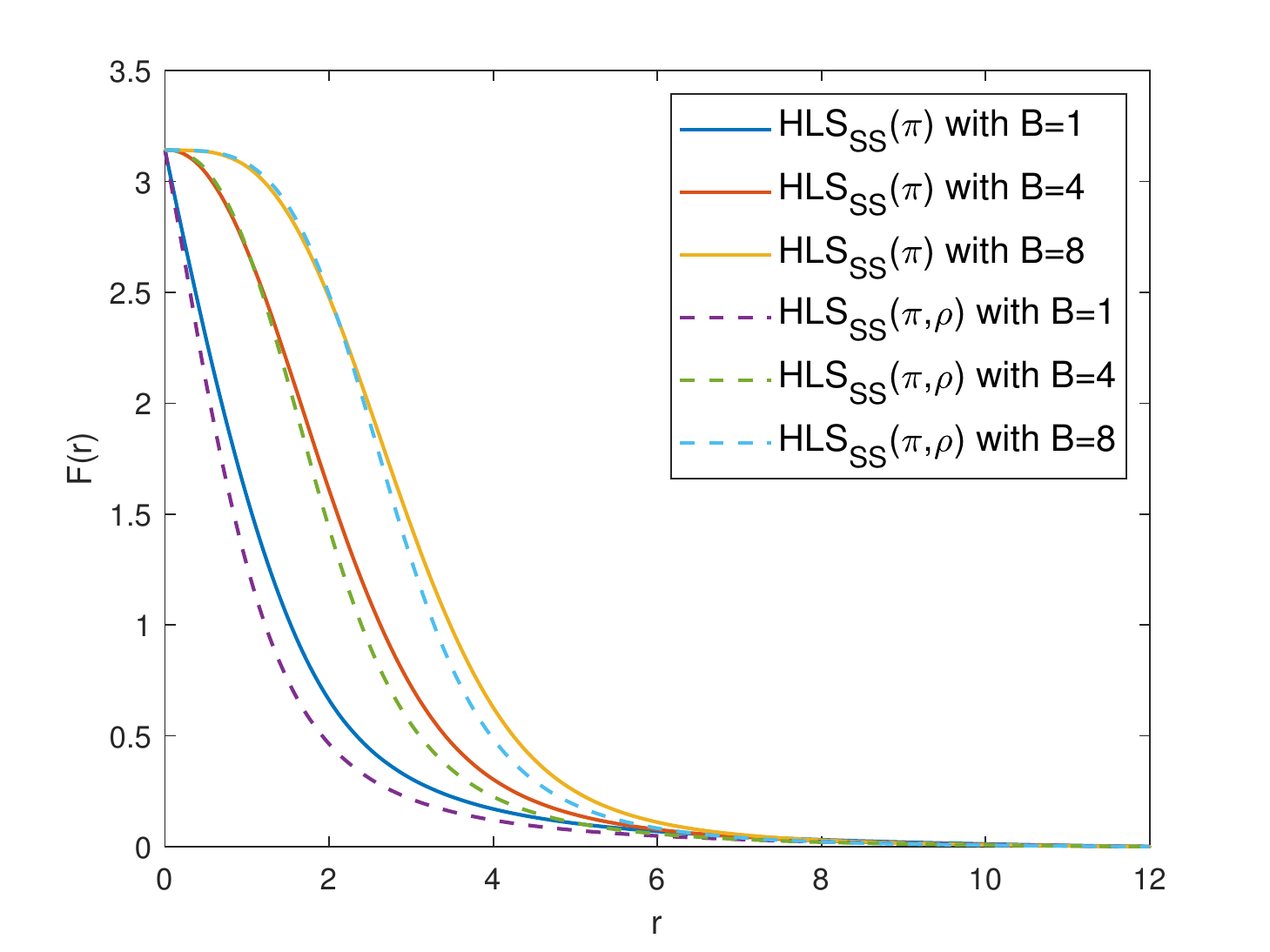}
	\caption{Profile function $F(r)$ in HLS$_{\rm SS}(\pi,\rho)$  and  HLS$_{\rm SS}(\pi)$ with $B=1,4,8$.}
	\label{fig:FSSPRPcompare}
\end{figure}

\begin{figure}[htbp]
	\centering
	\includegraphics[width=7.0cm]{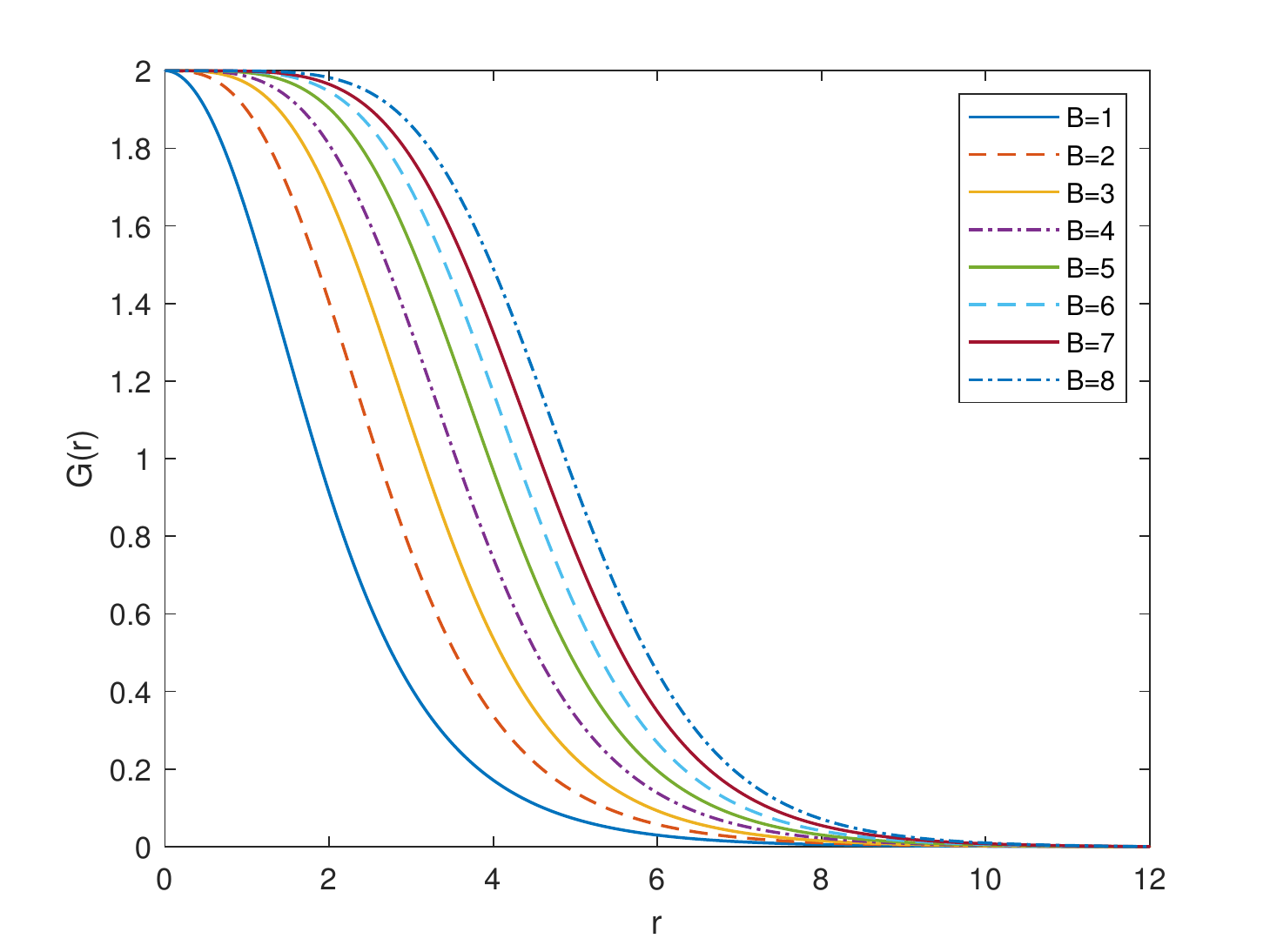}
	\includegraphics[width=7.0cm]{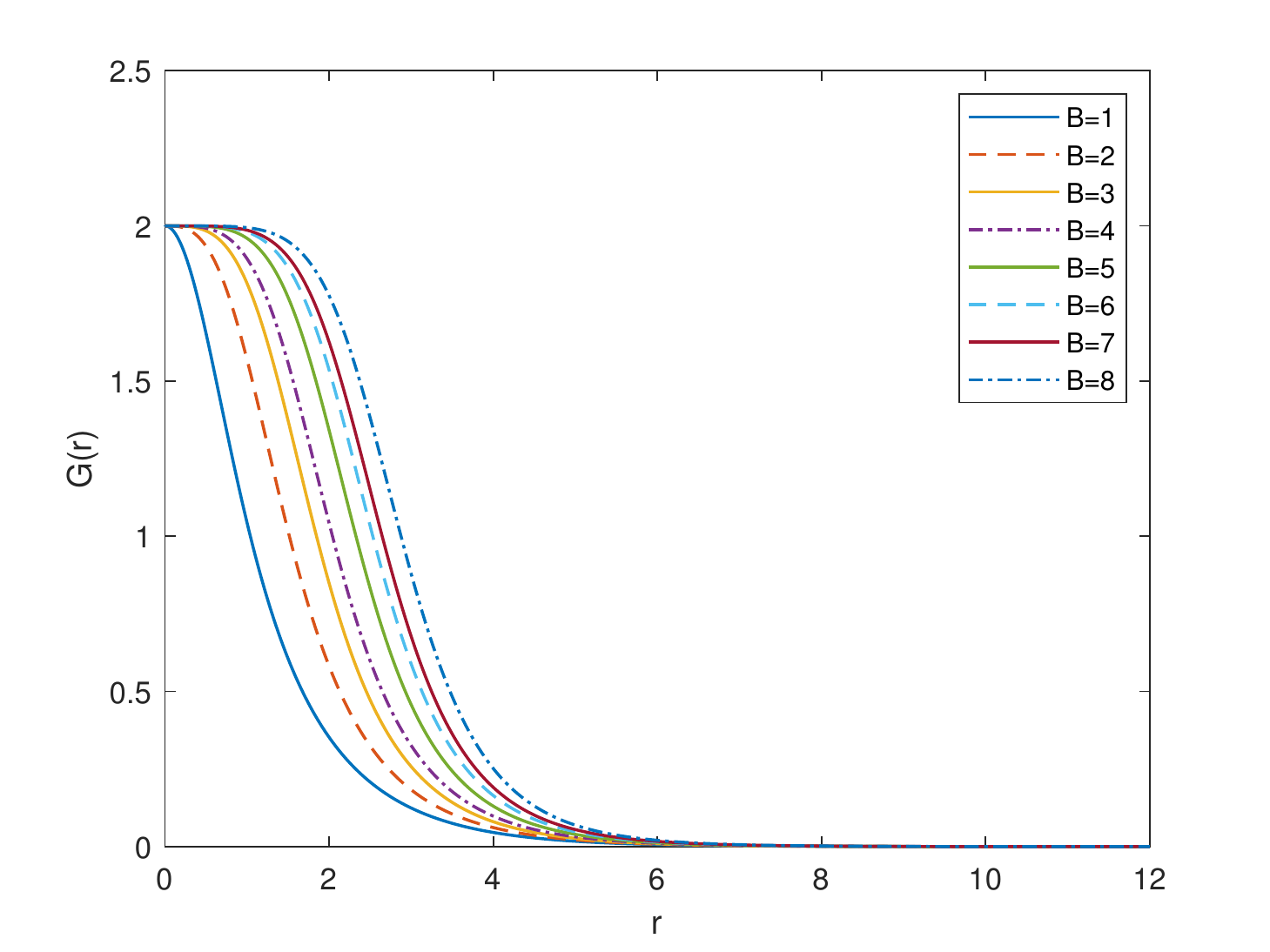}
	\caption{Profile function $G(r)$ in HLS$_{\rm SS}(\pi,\rho,\omega)$ (upper) and  HLS$_{\rm SS}(\pi,\rho)$ (lower).}
	\label{fig:GSS}
\end{figure}

We plot in Fig.~\ref{fig:FSS} the profile function $F(r)$ in the multi-skyrmion states. One can see that, due to the attraction from the rho meson, compared to HLS$_{\rm SS}(\pi)$, $F(r)$ shrinks in HLS$_{\rm SS}(\pi,\rho)$ a little bit, which can be explicitly seen in  Fig.~\ref{fig:FSSPRPcompare} for typical values of $B$. Different from the rho meson, the omega force in  HLS$_{\rm SS}(\pi,\rho,\omega)$ expands the distribution of $F(r)$ in a clear way. The same situation happens in the profile function $G(r)$ shown in Fig~\ref{fig:GSS}. From the expansions of the profile functions, we conclude that the size of the multi-skyrmion state from HLS$_{\rm SS}(\pi,\rho,\omega)$ is bigger than the corresponding ones from HLS$_{\rm SS}(\pi,\rho)$ and HLS$_{\rm SS}(\pi)$ and the corresponding state from  HLS$_{\rm SS}(\pi,\rho)$ has the smallest size.

\subsubsection{Effect from the warping factor}

We next study the effects of the warping factor by comparing the results calculated from HLS$_{\rm SS}$ and HLS$_{\rm BPS}$.  

We first list the masses of the multi-skyrmion states in Table~\ref{tab:MassBPS} for the baryon numbers $B=1,2,\cdots, 8$ using the LECs calculated from the BPS model. The results clearly show that, the same as the HLS$_{\rm SS}$, due to the attractive force from rho meson, the masses of the multi-skyrmion states from HLS$_{\rm BPS}(\pi,\rho)$ are smaller than the corresponding ones from HLS$_{\rm BPS}(\pi)$. However, the repulsive force from omega meson increases the masses very much. Similar to HLS$_{\rm SS}$, all the states in HLS$_{\rm BPS}(\pi)$ and HLS$_{\rm BPS}(\pi, \rho,\omega)$ are bound ones but all the states from  HLS$_{\rm BPS}(\pi,\rho)$ list here are not bound.

\begin{table}[htp]\centering
	\caption{Masses of the multi-skyrmion states in HLS$_{\rm BPS}$ (in unit of $ 4 \pi f_{\pi}^2/m_{\rho} $). Only the hadron degrees of freedom are explicitly written for simplicity.}
	\label{tab:MassBPS}
	\begin{tabular}{ccccccccc} \hline\hline
		B         & 1    & 2     & 3     & 4     & 5     & 6     & 7     & 8
		\\ \hline
		($\pi,\rho,\omega$)   & 8.40 & 16.74 & 24.86 & 32.64 & 40.73 & 48.61 & 56.17 & 64.25 \cr
		($\pi,\rho$)         & 4.17 & 8.75  & 13.16 & 16.93 & 21.58 & 25.79 & 29.25 & 33.94 \cr
		 ($\pi$)    & 4.87  & 9.54 & 14.03 & 17.94 & 22.64 & 26.94 & 30.59 & 35.30 \cr
		\hline\hline
	\end{tabular}
\end{table}

Comparing the results from HLS$_{\rm SS}$ and HLS$_{\rm BPS}$ we see that due to the warping factor in the SS model, the masses of the multi-skyrmion states calculated in the former are larger than the corresponding ones in the latter. In the HLS$_{\rm BPS}(\pi, \rho)$, the binding energy is very small which is because the BPS model is close to the BPS limit~\cite{Sutcliffe:2008sk} and therefore is very difficult to form bound states.

\begin{figure*}[htbp] \centering
    \includegraphics[width=1\linewidth]{fig_baryon_density_pr_ss}
    \includegraphics[width=1\linewidth]{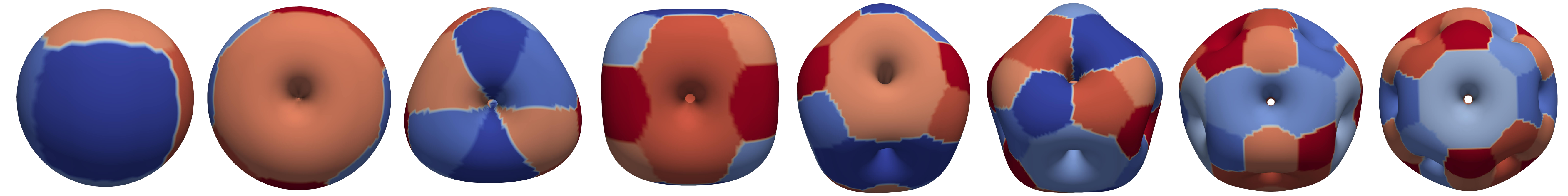}
	\caption{Contour surface with $\mathcal{B}^0=0.01$ in the $\text{HLS}_{\rm SS}(\pi,\rho)$(upper) and $\text{HLS}_{\rm BPS}(\pi,\rho)$ (lower) for $B=1,2, \cdots, 8$ (from left to right).} 
	\label{fig:contourprSSBPS}
\end{figure*}

\begin{figure}[htbp]
	\centering
	\includegraphics[width=8.0cm]{fig_ss_pr_F}
	\includegraphics[width=8.0cm]{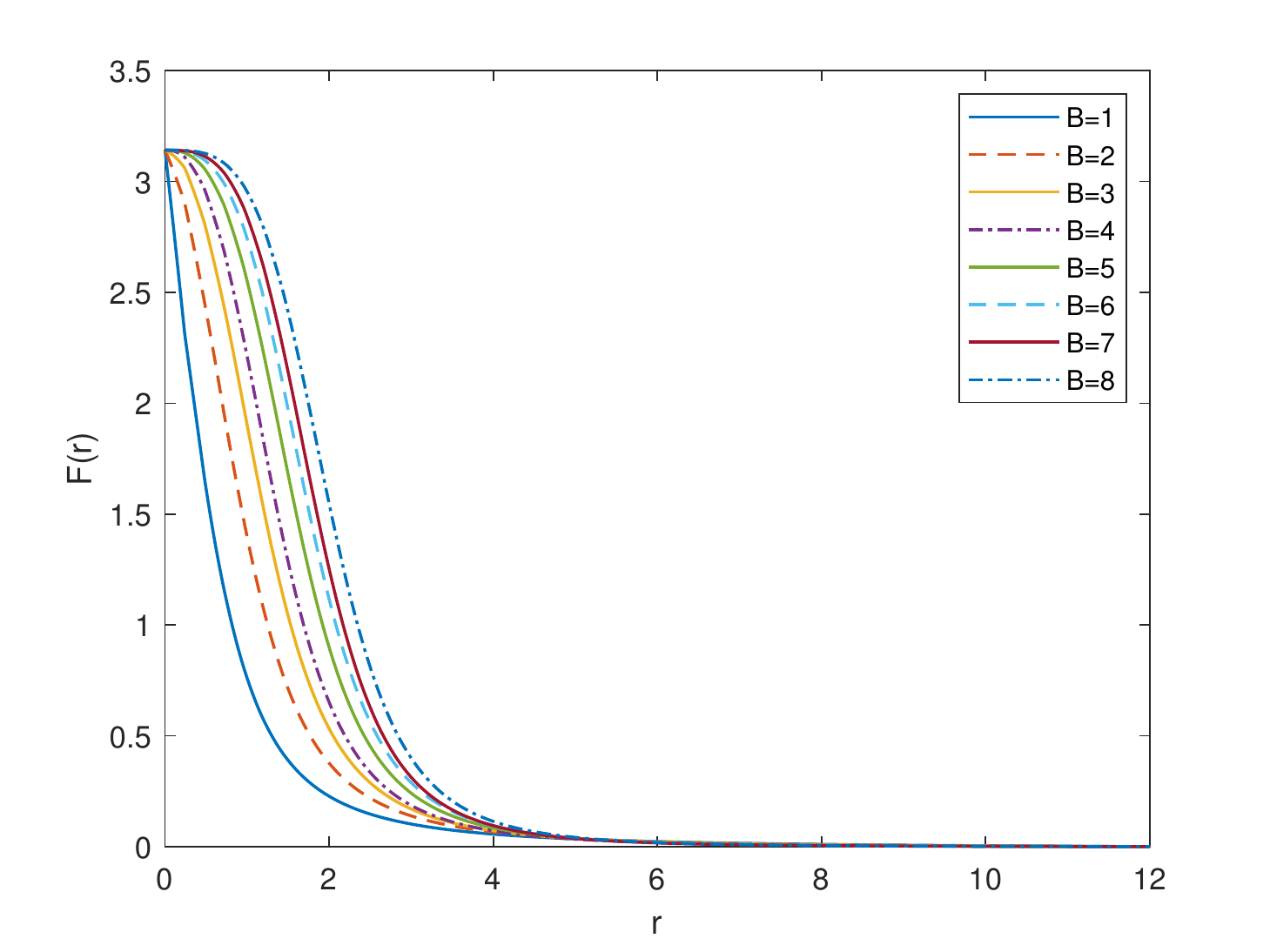}
	\caption{Profile functions $F(r)$ in HLS$_{\rm SS}(\pi,\rho)$ (upper) and HLS$_{\rm BPS}(\pi,\rho)$ (lower) models.}
	\label{fig:FprSSBPS}
\end{figure}

To further understand the effect of the warping factor, we compare the contour surface of the baryon number density. Here we take the results from HLS$_{\rm SS}(\pi,\rho)$ and HLS$_{\rm BPS}(\pi,\rho)$ as examples and plot the results for the baryon number density $\mathcal{B}^0=0.01$ in Fig.~\ref{fig:contourprSSBPS}. This figure explicitly shows that, due to the warping factor in the SS model, the constituents of the multi-skyrmion states are far away from each other. Since the constituents are further, the distributions of the profile functions are more expanded in HLS$_{\rm SS}$ as shown in Fig.~\ref{fig:FprSSBPS}.

\section{Summary and discussion}

\label{sec:conclusion} 

Using the hidden local symmetry approach, we introduce the vector mesons $\rho$ and $\omega$ into the Skyrme model and calculate the effect of vector mesons on the multi-skyrmion states with baryon numbers from $B=1$ to $8$. With the help of the holographic models, all the LECs can be self-consistently calculated by using two inputs $f_\pi$ and $m_\rho$. In this sense, we explicitly studied the effects of the vector mesons and the warping factor on the properties of multi-skyrmion states.

The main conclusions of this work can be summarized as follows: 
Compared to the model with pion only, the $\rho$ meson slightly reduces the mass of the multi-skyrmion states, and the profile function $F(r)$ also slightly shrinks. The effect of the $\omega$ meson on the multi-skyrmion states is obvious, not only on increases the masses of the multi-skyrmion states but also expansions the sizes of the states; The contribution from the warping factor cannot be ignored if the LECs of the HLS are estimated with the help of the holographic models of QCD.

Given the above qualitative conclusions, several extensions of the present work can be expected.

In this work, we calculated the LECs using the SS model and the BPS model. It is well known that the hadron spectrum, including that of the skyrmions, does not agree with the empirical values. To avoid this defect, it is interesting to resort to certain holographic models which can yield hadron properties consistent with nature. A possible approach is to use the holographic model from the bottom-up approach, for example, the soft-wall model developed in Refs.~\cite{Cui:2013xva,Fang:2019lmd}, making use of the approach developed in Ref.~\cite{Harada:2014lza}. In addition, in this approach, the explicit chiral symmetry breaking effect which is ignored in the present work but is found significant for the spectrum and shapes of the multi-skyrmion states can be self-consistently taken into account.

A generic problem in the skyrmion approach to nuclear physics is that both the masses of the baryons and binding energies between skyrmions are too large to confront the empirical values. It is promising to overcome these problems in some modified Skyrme models such as the BPS Skyrme model~\cite{Adam:2013wya} and the false vacuum model~\cite{Ferreira:2021ryf}. So it is interesting to check the properties of the multi-skyrmion states using these models.

Generally speaking, the isospin unit vector $\hat{\bm{n}}$ in the parametrization of the rho meson~\eqref{eq:ansatzV} should be different from that in the parametrization of the pion field~\eqref{eq:ansatzpion}. In other words, there should be two different vectors, $\hat{\bm{n}}^{\rho}$ for the rho field and $\hat{\bm{n}}^{\pi}$ for the pion field, since they are independent fields, like what has been done in the skyrmion crystal approach to nuclear matter~\cite{Park:2003sd}. We actually checked such a senario that $\hat{\bm{n}}^{\pi}$ and $\hat{\bm{n}}^{\rho}$ are differently parameterized. We finally found that to yield the minimal mass of the multi-skyrmion state, the same form of the paramererization has to be used, as what was done above.

For the purpose to see the effects of the vector mesons and warping factor, we focus on the rational map ansatz here. For some specific models and multi-skyrmion states, the rational map ansatz does not always yield the global minima. To confirm the real structure of the multi-skyrmion states, not only the rational map ansatz but also the product ansatz should be equally checked.

\acknowledgments

We would like to thank Y. Tian and H. B. Zhang for their valuable discussions. The work of Y.~L. M. was supported in part by the National Science Foundation of China (NSFC) under Grant No. 11875147 and No. 12147103.

\appendix

\section{Mass of the multi-skyrmion state in the hidden local symmetry}

\label{sec:app}

In this appendix, we provide the expression of the mass of the multi-skyrmion state using the Lagrangian~\eqref{eq:HLSall}. In accordance with this Lagrangian, we decompose the mass as
\be
M & = & M_{(2)} + \sum_{i=1}^9 y_i \frac{1}{2} ag^2 M_{y_i} + z_4 ag^2 m_{z_4} + z_5 ag^2 M_{z_5} \nonumber\\
& &{} + \sum_{i=1}^3 c_i \frac{ ag^3 N_c}{16 \pi^2} M_{c_i}.
\ee
In unit of the scale factor $4 \pi \frac{f_{\pi}^2}{m_{\rho}}$, we explicitly have
\begin{widetext}
\be
M_{(2)} & = & \int dr \left \{ \frac{1}{2}r^2 F^{\prime 2} + n\sin ^2 F-\frac{ag^2}{2}r^2 W^2 
    -\frac{ag^2}{2}r^2 W^{\prime 2} + an(G-1+\cos F)^2 
    +anG^{\prime 2}+\frac{aI}{2r^2}G^2(G-2)^2 \right \}\nonumber\\
%%%%%%%%%%%%%%%%%%%%%%%%%%%%%%%%%%%%%%%%%%%%%%%%%%%%%%%%%%%%%%%%%%%%%%%%%%%%%%%%%%%%%%%%
M_{y_1} & = &{} - \int dr \left( \frac{1}{4} r^2F^{\prime 4} +n F^{\prime 2} \sin ^2 F + \frac{I}{r^2}\sin ^4 F \right), \nonumber\\
%%%%%%%%%%%%%%%%%%%%%%%%%%%%%%%%%%%%%%%%%%%%%%%%%%%%%%%%%%%%%%%%%%%%%%%%%%%%%%%%%%%%%%%%
M_{y_2} & = &{} - \int \,dr \left( \frac{1}{4} r^2F^{\prime 4} -n F^{\prime 2} \sin ^2 F \right), \nonumber\\
%%%%%%%%%%%%%%%%%%%%%%%%%%%%%%%%%%%%%%%%%%%%%%%%%%%%%%%%%%%%%%%%%%%%%%%%%%%%%%%%%%%%%%%%
M_{y_3} & = &  \int \,dr \left( -\frac{g^4}{4} r^2 W^4 + g^2 W^2 n (G-1+\cos F)^2 - \frac{I}{r^2}(G-1+\cos F)^4 \right), \nonumber\\
%%%%%%%%%%%%%%%%%%%%%%%%%%%%%%%%%%%%%%%%%%%%%%%%%%%%%%%%%%%%%%%%%%%%%%%%%%%%%%%%%%%%%%%%
M_{y_4} & = &  \int \,dr\left( -\frac{g^4}{4} r^2 W^4 + g^2 W^2 n (G-1+\cos F)^2  \right), \nonumber\\
%%%%%%%%%%%%%%%%%%%%%%%%%%%%%%%%%%%%%%%%%%%%%%%%%%%%%%%%%%%%%%%%%%%%%%%%%%%%%%%%%%%%%%%%
M_{y_5} & = &  \int \,dr \left( \frac{g^2}{4} r^2 W^2 F^{\prime 2}- \frac{1}{2} n F^{\prime 2} (G-1+\cos F)^2 + \frac{1}{2} g^2 n W^2 \sin^2F - \frac{I}{r^2} \sin^2F (G-1+\cos F)^2 \right), \nonumber\\
%%%%%%%%%%%%%%%%%%%%%%%%%%%%%%%%%%%%%%%%%%%%%%%%%%%%%%%%%%%%%%%%%%%%%%%%%%%%%%%%%%%%%%%%
M_{y_6} & = & 0, \nonumber\\
%%%%%%%%%%%%%%%%%%%%%%%%%%%%%%%%%%%%%%%%%%%%%%%%%%%%%%%%%%%%%%%%%%%%%%%%%%%%%%%%%%%%%%%%
M_{y_7} & = &{} - \int \,dr \frac{I}{r^2} \sin^2F (G-1+\cos F)^2, \nonumber\\
%%%%%%%%%%%%%%%%%%%%%%%%%%%%%%%%%%%%%%%%%%%%%%%%%%%%%%%%%%%%%%%%%%%%%%%%%%%%%%%%%%%%%%%%
M_{y_8} & = & 2 \int \,dr \frac{I}{r^2} \sin^2F (G-1+\cos F)^2, \nonumber\\
%%%%%%%%%%%%%%%%%%%%%%%%%%%%%%%%%%%%%%%%%%%%%%%%%%%%%%%%%%%%%%%%%%%%%%%%%%%%%%%%%%%%%%%%
M_{y_9} & = & \frac{1}{2} \int \,dr\left( \frac{1}{2} g^2 W^2 F^{\prime 2} r^2 + g^2 W^2 n \sin^2F + F^{\prime 2} n (G-1+\cos F)^2 \right), \nonumber\\
%%%%%%%%%%%%%%%%%%%%%%%%%%%%%%%%%%%%%%%%%%%%%%%%%%%%%%%%%%%%%%%%%%%%%%%%%%%%%%%%%%%%%%%%
M_{z_4} & = & \int \,dr \left(-n F^\prime G^\prime \sin F + \frac{I}{2r^2}\sin^2F G(G-2) \right), \nonumber\\
%%%%%%%%%%%%%%%%%%%%%%%%%%%%%%%%%%%%%%%%%%%%%%%%%%%%%%%%%%%%%%%%%%%%%%%%%%%%%%%%%%%%%%%%
M_{z_5} & = & \int \,dr \frac{I}{2r^2} (G-1+\cos F)^2 G(G-2), \nonumber\\
%%%%%%%%%%%%%%%%%%%%%%%%%%%%%%%%%%%%%%%%%%%%%%%%%%%%%%%%%%%%%%%%%%%%%%%%%%%%%%%%%%%%%%%%
M_{c_1} & = &  \int \,dr \left( n W F^\prime(G-1+\cos F)^2 + 3nWF^\prime\sin ^2F \right), \nonumber\\
%%%%%%%%%%%%%%%%%%%%%%%%%%%%%%%%%%%%%%%%%%%%%%%%%%%%%%%%%%%%%%%%%%%%%%%%%%%%%%%%%%%%%%%%
M_{c_2} & = & \int \,dr \left( n W F^\prime(G-1+\cos F)^2 - 3nWF^\prime\sin ^2F \right), \nonumber\\
%%%%%%%%%%%%%%%%%%%%%%%%%%%%%%%%%%%%%%%%%%%%%%%%%%%%%%%%%%%%%%%%%%%%%%%%%%%%%%%%%%%%%%%%
M_{c_3} & = & 2  \int \,dr \left( n W \sin F G^\prime - n W^\prime \sin F (G-1+\cos F) - \frac{1}{2} n W F^\prime G(G-2) \right),
\ee
where baryon number $n$ and function $I$ are defined as
\be
n & = & \frac{1}{4\pi} \int \left(\frac{1+\left\lvert z\right\rvert ^2 }{1+\left\lvert R\right\rvert ^2} \left\lvert \frac{dR}{dz} \right\rvert \right)^2 \frac{ 2i dz d \bar{z} }{(1+\left\lvert z\right\rvert ^2)^2} \\
I & = & \frac{1}{4\pi} \int \left(\frac{1+\left\lvert z\right\rvert ^2 }{1+\left\lvert R\right\rvert ^2} \left\lvert \frac{dR}{dz} \right\rvert \right)^4 \frac{ 2i dz d \bar{z} }{(1+\left\lvert z\right\rvert ^2)^2}
    \label{baryon_num_and_I}
\ee

\end{widetext}

\bibliography{SkyrNucleiHLSRef}

\end{document}